\documentclass{article}
\pdfoutput=1
\usepackage{amsmath, amssymb}
\usepackage{graphicx}
\usepackage{multicol}
\usepackage{pstricks}
\usepackage{color}
\title{Position Coding}
\author{Edward Aboufadel (Grand Valley State University) \\ Liz Smietana (John Carroll University)
 \\ Tim Armstrong (Reed College)}
\date{June 1, 2007 draft}
\begin{document}
\maketitle

\section{The Fly Pentop Computer}
Advertised as ``a computer in a pen'', the Fly Pentop computer (ref web site)  hit the
stores in the fall of 2005.  One application of the pen is a four-function
calculator that the user \emph{draws} on a special kind of paper called ``Fly
paper''.  The pen seems to know what numbers have been written and where on the paper the numbers lie,
because after tapping on what was drawn (e.g. 5, +, 3, 7, =), the computer
announces the result of ``42''.  The pen can also read handwriting,
and there are a number of games printed on ``Fly paper'' that are
based on the pen knowing the location of objects on the paper.  How
does the pen know where symbols and objects are on the paper?  And
what does the following quote from a recent article (cite Wired) about
this computer mean:  ``The pen knows where it is within a
1.8-million-square-mile grid''?

The Fly Pentop Computer uses paper which has a \emph{position code}
lightly printed on it.  A position coding pattern is an array of
symbols in which subarrays of a certain fixed size appear at most
once.  So, each subarray uniquely identifies a location in the larger
array, which means  there is a bijection of some sort from this set of subarrays
to a set of coordinates.  The key to ``Fly paper'' and other examples
of position codes is a method to read the subarray and then convert it
to coordinates.  So, for the calculator application, when the
calculator is first drawn,  the pen reads a symbol such as ``2''
through optical character recognition, and also
reads the position of the ``2'' by interpreting the position code on
the paper.  When the ``2'' is later tapped, the position code is read
again, and the fact that a ``2'' is there is recovered from the pen's
memory.

Position coding makes use of ideas from discrete mathematics and
number theory.  In this paper, we will describe the underlying
mathematics of two position codes, one being the Anoto code that is
the basis of ``Fly paper''.  Then, we will present two new codes, one
which uses binary wavelets as part of the bijection.

\section{The Resnik Position Code}

The CCD Rasnik Straightness Monitoring System~\cite{Rasnik} is a position coding pattern that was designed for the muon detector at CERN, the European Organization for Nuclear Research.  Because the detector could not be built to be sufficiently stable to measure absolute positions, the relative position of the components are continously monitored.  This is done by determining the positions of lightspots shined on the components.

\begin{figure}[!h]
\begin{center}
\includegraphics[width=2in]{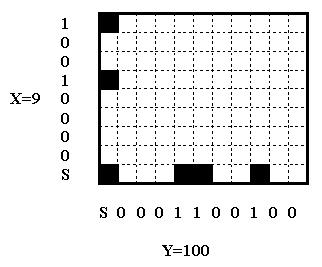}
\caption{Example of a B3}
\label{b3}
\end{center}
\end{figure}

In this pattern, each position coding block, which is called a B3,
codes for a unique position.  Each B3 is composed of smaller pixels,
and the size of a B3 is $9\times11$ pixels.  The coding for the $x$
coordinate occurs in the first column of a B3, and the coding for the
$y$ coordinate occurs in the bottom row.  This coding is done by using
codebits, which appear as black squares.  If a codebit is present in a
particular field, that field assumes the value 1.  All fields without
codebits are considered 0.  To determine the $x$ coordinate, one
simply translates the first column into 0s and 1s, and the binary
number associated with this column is the appropriate $x$ coordinate.
The same is true for determining the $y$ coordinate along the bottom
row.  Figure \ref{b3} shows an example of a B3.  Coordinates in the
range $0 \le x \le 255$, $0 
\le y \le 1023$ can be coded.  In addition, each B3 has a ``startbit'' in the lower left-hand corner, for reference.

\begin{figure}[!h]
\begin{center}
\includegraphics[width=2in]{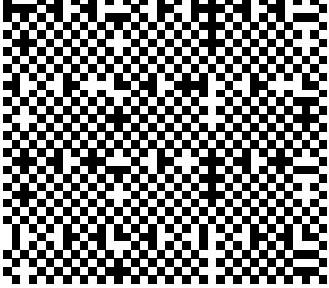}
\caption{Position Coding Pattern for the Rasnik Mask}
\label{rasnik}
\end{center}
\end{figure}

In order to make the this position coding appear more random, and to create a more stable average illumination of the pattern,  each B3 is overlayed with a $9\times11$ chessboard.  If a codebit is present, the particular chess square associated with this codebit will be converted to the opposite color.  That is, if a codebit is overlayed with black square from a chessboard, that square will be converted to white, and if the square was white, it would be changed to black.  Figure \ref{rasnik} shows a sample of this position coding pattern, which contains several B3s.

To determine a Resnik location, a $9\times11$ subarray is read, and
the chessboard overlay is undone via XOR, yielding a B3 as in Figure 1, or
part of a B3 (which we will call the ``main B3'') along with parts of
adjacent B3s.  Because adjacent B3s differ in their $x$ or $y$
coordinates by one unit, because of the reference startbits, and because the bottom row and left column
are all that are neeed to determine the postion,
the unknown parts of the main B3 can be deduced from the information
in the adjacent B3s.

\section{The Anoto Pattern}
The position coding pattern used by ``Fly Paper'' is the Anoto
pattern, patented by the Anoto corporation. It appears that the precise pattern found in patent WO 03/038741~\cite{anoto} is used by ``Fly paper'', based on an analysis of samples.  This came as a bit of surprise, as Anoto describes the details in this patent as merely an ``example''.  

\begin{figure}
\begin{center}
\includegraphics{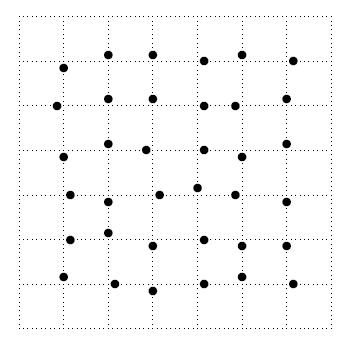}
\caption{Fly Paper}
\label{flypaper}
\end{center}
\end{figure}

 The Anoto pattern is made up of many small dots offset from a grid,
 which is called a raster.  Each of the dots is offset in one of four
 directions, thus encoding two bits of information.  A diagram of an
 enlarged piece of the pattern is shown in figure \ref{flypaper}. This diagram is a typical subarray that can be associated with an $(x, y)$ pair.   The dotted lines in this diagram represent the grid, which does not actually appear on the paper.  The pattern has the property that any $6\times 6$ array of dots appears only once.  This allows a device to determine a location within the grid by only looking at a $6\times 6$ array.

A certain type of sequence, called a De Bruijn sequence, is related to some aspects of the code.  Given a finite set $S=\{s_1,s_2,\ldots,s_k\}$, a cyclic sequence of length $k^n$ of elements in $S$ in which every subsequence of length $n$ appears exactly once is called a De Bruijn sequence of order $n$~\cite{Wolf}.  Some of the sequences in the pattern are similar to De Bruijn sequences, but are not of maximal length.  We will call a cyclic sequence of length $m<k^n$ of elements of $S=\{s_1,s_2,\ldots,s_k\}$ in which any subsequence of length $n$ appears no more than once a \emph{quasi De Bruijn} sequence of length $m$ from the alphabet $S$.  The problem of finding quasi De Bruijn sequences is related to the \emph{shortest common superstring problem}~\cite{Black}.

In the pattern, one bit from each dot is used to code the $x$ coordinate and the other is used to code the $y$ coordinate.  For each coordinate, there is a repeated quasi De Bruijn sequence of order $6$ of length $63$ from the alphabet $\{0,1\}$.  This sequence is called the main number sequence (for a given coordinate).  The main number sequence used in the Anoto pattern is the following:
\begin{align*}
&\{0,0,0,0,0,0,1,0,0,1,1,1,1,1,0,1,0,0,1,0,0,0,0,1,1,1,0,1,1,1,0,0,1,0, 1,0,  \\
&1,0,0,0,1,0,1,1,0,1,1,0,0,1,1,0,1,0,1,1,1,1,0,0,0,1,1\}.
\end{align*}

\begin{figure}[!h]
\begin{center}
\includegraphics{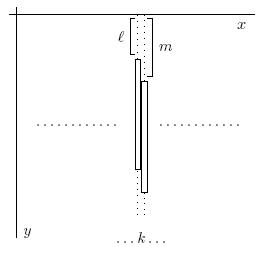}
\caption{Calculation of a Member of the Primary Difference Sequence}
\label{mainseq}
\end{center}
\end{figure}

For the $x$ direction, translations of the main number sequence by numbers between $0$ and $62$ are coded in columns.  That is, each column contains the same cyclic sequence moved up or down by some number of rows.  The difference between the amount by which any two adjacent columns are translated forms a member of a sequence called the primary difference sequence. We will denote this sequence by $\{d_n\}_{n\ge 1}$.  In other words, if the $k^\text{th}$ column is translated down by $\ell$ and the $k+1^\text{st}$ column is translated down by $m$, then $d_k=\ell-m \pmod{63}$.  This is shown in Figure \ref{mainseq}.  Since the main number sequence is a quasi De Bruijn sequence of order $6$, it is possible to determine $d_k$ using adjacent subsequences of length $6$ in the $k^\text{th}$ and $k+1^\text{st}$ columns.  If the subsequence from column $k$ starts at $\ell$ and the adjacent subsequence from column $k+1$ starts at $m$, then $d_k=\ell-m \pmod{63}$.

Since the members of the primary difference sequence are determined modulo $63$, each member must be between $0$ and $62$.  However, as stated in the patent, only numbers in the set $\{5,\ldots,58\}$ are allowed.  This set has $54=2\cdot 3^3$ elements, so there are bijections from $\{5,\ldots,58\}$ to $\{(a_1,a_2,a_3,a_4)|a_1,a_2,a_4\in \{0,1,2\}, a_3\in\{0,1\}\}$.  We can define one such bijection, which we will call $\phi$, by the following rule.  For any $r\in \{5,\ldots,58\}$, there is a unique $(a_1,a_2,a_3,a_4)\in\{(a_1,a_2,a_3,a_4)|a_1,a_2,a_4\in \{0,1,2\}, a_3\in\{0,1\}\}$ such that 
\[
r=5+a_1+a_2\cdot 3+a_3\cdot 3^2+a_4\cdot 2\cdot3^2
\]
Let $\phi(r)=(\phi_1(r),\phi_2(r),\phi_3(r),\phi_4(r))=(a_1,a_2,a_3,a_4)$.

For $\ell$ from $1$ to $4$, define the sequence $\{a_{\ell,n}\}_{n\ge 1}$ by $a_{\ell,k}=\phi_\ell(d_k)$.  These sequences are called the secondary number sequences. Each of the sequences $\{a_{1,n}\}_{n\ge 1}$, $\{a_{2,n}\}_{n\ge 1}$, and $\{a_{4,n}\}_{n\ge 1}$ are formed by repeating quasi De Bruijn sequences from the alphabet $\{0,1,2\}$ of order five of lengths $236$, $233$, and $241$ respectively, while the sequence $\{a_{3,n}\}_{n\ge 1}$ is formed by repeating a quasi De Bruijn sequence from the alphabet $\{0,1\}$ of order five of length $31$.  The lengths of these quasi De Bruijn sequences are chosen to be relatively prime so that the least common multiple of all of the lengths is their product.  For Fly Paper, $236$, $233$, $31$ and $241$ are relatively prime, so $\text{lcm}(236,233,31,241)=236\cdot 233\cdot 31\cdot 241=410815348$.  So, for any $d_k$ in the primary difference sequence, $d_k=\phi^{-1}((a_{1,k},a_{2,k},a_{3,k},a_{4,k}))$, which means that any subsequence of length $5$ in the primary difference sequence occurs once every $410815348$ elements.

The position of $d_k$ in the primary difference sequence is one of the pieces of information that codes the $x$ coordinate.  However, the same primary difference sequence can be achieved if the translation of the main number sequence in each column is translated down by $\ell$ rows for any $\ell$ from $0$ to $62$.  How far down all of the translations of the main number sequences are translated determines another piece of information about the $x$ coordinate, called the section.  Thus, the $x$ coordinate is encoded by a section and a place in the primary difference sequence.  The encoding of the $y$ coordinate uses the remaining bit and is similar to the encoding of the $x$ coordinate with all of the sequences perpendicular to their counterparts for the $x$ coordinate.  Both coordinates can use the same system of sequences.

We have written a \emph{Maple} program, which can be found in the \emph{Maple} worksheet \verb/anoto_work2.mws/, to convert a scanned image of the Anoto pattern into arrays of bits and look for the sequences described in the patent.  Using this \emph{Maple} program, we have found the main number sequence and secondary difference sequences for encoding the $y$ coordinate that seem to be the same as the ones described in the patent.  We have also found the main number sequence and the base $2$ secondary difference sequence for the $x$ direction.  These sequences are identical to the ones used for the $y$ coordinate.  Although we have not found the other secondary difference sequences for the $x$ coordinate, we expect that they are the same.

\section{deBrujin tori}

One topic that may be useful in creating position coding patterns is the study of De Bruijn tori.  An $(R,S;m,n)_k$-De Bruijn torus is a two dimensional toroidal array with $R$ rows and $S$ columns in which every $m\times n$ array of elements from a set with $k$ elements appears exactly once, including the parts of the torus where the array wraps around and meets itself~\cite{HandI}.  Such an array satisfies the necessary property for position coding patterns of having every $m\times n$ array appear at most once.  However, a De Bruijn torus may not be an ideal position coding pattern if there is no efficient way of finding the position of a given section.

\section{A Position Coding Pattern that Uses Binary Wavelets}

We have created two position coding patterns, one of which uses binary wavelets.  Binary field wavelets are quite similar to wavelets of the real field.  Binary wavelet filters are constructed to satisfy certain properties, some of which are analagous to the real field wavelets.  Three key properties that are important are the bandwidth constraint, the vanishing moments constraint, and the perfect reconstruction constraint~\cite{Binary}.  The perfect reconstruction constraint states that the wavelet transform must be invertible, while the others ensure that wavelet transforms of certain arrays have certain desirable properties.  For example, the vanishing moments constraint ensures that the transforms of nearly linear arrays have few nonzero entries in the detail filters.  

We have created a position coding pattern using these binary wavelet filters.  The filter that we are using is determined by the matrix
\[
T=
\begin{bmatrix}
1&0&1&1\\
1&1&1&0\\
1&1&0&0\\
0&0&1&1
\end{bmatrix}
\]
The transform of a $4\times4$ matrix $F$ is given by $TFT^t$.  For the coding of our binary wavelet pattern, each coordinate is determined by a $4\times4$ matrix.  Specifically, the top two rows of each matrix determine the $x$ coordinate and the bottom two rows determine the $y$ coordinate.  This coding is done simply by using binary numbers, as shown in the matrix below.  The position values in blue (the top two rows) represent the $x$ coordinate, and the values in red (the bottom two rows) represent the $y$ coordinate.
\[
\begin{bmatrix}
\textcolor{blue}{2^0}&\textcolor{blue}{2^1}&\textcolor{blue}{2^2}&\textcolor{blue}{2^3}\\
\textcolor{blue}{2^4}&\textcolor{blue}{2^5}&\textcolor{blue}{2^6}&\textcolor{blue}{2^7}\\
\textcolor{red}{2^0}&\textcolor{red}{2^1}&\textcolor{red}{2^2}&\textcolor{red}{2^3}\\
\textcolor{red}{2^4}&\textcolor{red}{2^5}&\textcolor{red}{2^6}&\textcolor{red}{2^7}\\\end{bmatrix}
\] 
As an example, suppose we wished to code for the position $\left(12,108\right)$.  Then the corresponding matrix $G$ would appear as follows:
\[
G=
\begin{bmatrix}
0&0&1&1\\
0&0&0&0\\
0&0&1&1\\
0&1&1&0
\end{bmatrix}
\]
To continue with the coding process, we would then apply the inverse of the binary wavelet transform to this matrix.  That is, we would compute $T^{-1}G(T^{t})^{-1}$ to attain a new matrix representation of the coordinate $\left(12,108\right)$.  This new representation is how the coordinate would appear in the position coding pattern.  

\begin{figure}[!h]
\begin{center}
\includegraphics{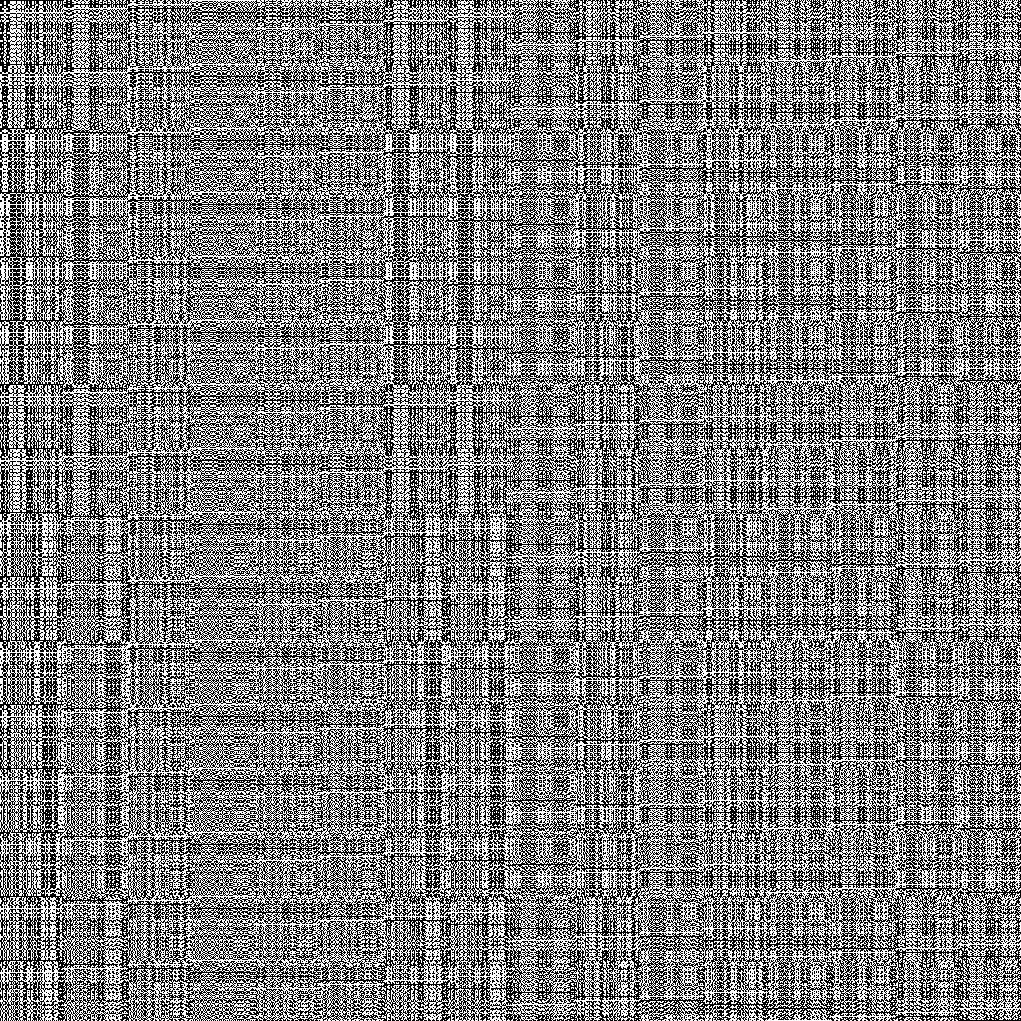}
\caption{Image of the Binary Position Coding Pattern}
\label{binary}
\end{center}
\end{figure}

\begin{figure}[!h]
\begin{center}
\includegraphics[width=2.5in]{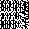}
\caption{Close-up of the Binary Position Coding Pattern}
\label{binary2}
\end{center}
\end{figure}

This binary wavelet pattern can code for positions $\left(0,0\right)$ through $\left(255,255\right)$, and so it can code of a total of $255\times255=65536$ unique positions.  Because each position is represented by a $4\times4$ matrix, the position coding pattern appears as a $1024\times1024$ binary matrix.  Additionally, we can represent this data as an image consisting of black and white pixels, where the black pixels stand for zeros and the white pixels stand for ones.  Figure \ref{binary} shows the entire position coding pattern as a black and white image, and figure \ref{binary2} is a close-up of a small portion of the image.     

While this position coding pattern works fairly well, its main limitation is that it requires gridlines or some form of delimiters in order to distinguish individual matrices.  One of our goals is to improve upon this and to eliminate the need for any form of delimiters.

\section{A Non Wavelet-Based Position Coding Pattern}
Although the wavelet based pattern relies on delimiters, we have created a method of making binary position coding patterns that does not use delimiters.  Given even $m$ and $n$, we can create a binary position coding pattern in which $m\times n$ sections are unique.

\subsection{The 4 by 4 Version of the Pattern}
\begin{figure}[!h]
\begin{center}
\includegraphics[width=4in]{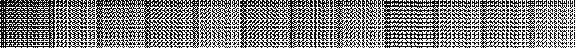}
\caption{Image of the Non Wavelet-Based Binary Position Coding Pattern}
\label{nwb}
\end{center}
\end{figure}

The 4 by 4 version of the pattern is a $48\times 576$ array in which any $4\times 4$ section appears at most once.  Each entry in the array is either 0 or 1.  The position of any $4\times 4$ section can be determined using a lookup table that is much smaller than the size of the actual pattern.  Figure \ref{nwb} shows the entire pattern.  A black pixel represents a zero, while a white pixel represents a one.  We will call the entire pattern $P$ and denote the entry $i-1$ units down and $j-1$ units to the right from the upper left entry by $P_{i,j}$ and call this entry the $i,j$th entry.

In any $4\times 4$ section, eight entries code two base 12 digits for the $x$ coordinate, four entries code a single base 12 digit for the $y$ coordinate, and the remaining four digits code the position of each coordinate modulo 4.  The pattern is created by meshing together the array
\[
U=\left[\begin{array}{cc}
1 & 0 \\
0 & 0
\end{array}\right]
\]
with arrays from the sequence
\begin{align*}
A=&\left[\begin{array}{cc}
0 & 0 \\
0 & 0
\end{array}\right],
\left[\begin{array}{cc}
0 & 1 \\
0 & 1
\end{array}\right],
\left[\begin{array}{cc}
0 & 0 \\
1 & 1
\end{array}\right],
\left[\begin{array}{cc}
0 & 1 \\
1 & 0
\end{array}\right],
\left[\begin{array}{cc}
0 & 1 \\
1 & 1
\end{array}\right],
\left[\begin{array}{cc}
1 & 0 \\
0 & 1
\end{array}\right],  \\
&\left[\begin{array}{cc}
1 & 0 \\
1 & 0
\end{array}\right]
\left[\begin{array}{cc}
1 & 0 \\
1 & 1
\end{array}\right],
\left[\begin{array}{cc}
1 & 1 \\
0 & 0
\end{array}\right],
\left[\begin{array}{cc}
1 & 1 \\
0 & 1
\end{array}\right],
\left[\begin{array}{cc}
1 & 1 \\
1 & 0
\end{array}\right],
\left[\begin{array}{cc}
1 & 1\\
1 & 1
\end{array}\right].
\end{align*}
The sequence $A$ is used to code base-12 digits for each $4\times 4$ block for which both coordinates for the upper left corner are congruent to 1 modulo 4.  Specifically, if $j=12(j_1-1) + (j_0-1)$ for some $j_0$ and $j_1$ from 1 to 12, then the $4\times 4$ array starting at $4i + 1,4j + 1$ is given by
\[
\left[\begin{array}{cccc}
1           & (A_{j_1})_{1,1} & 0           & (A_{j_1})_{1,2}  \\
(A_i)_{1,1} & (A_{j_0})_{1,1} & (A_i)_{2,1} & (A_{j_0})_{1,2}  \\
0           & (A_{j_1})_{2,1} & 0           & (A_{j_1})_{2,2}  \\
(A_i)_{1,2} & (A_{j_0})_{2,1} & (A_i)_{2,2} & (A_{j_0})_{2,2}
\end{array}\right]
\]
where $i$ and $j$ are nonnegative integers with $i\leq 11$ and $j\leq 143$.  Note that if $i$ and $j$ are both odd, $P_{i,j}$ is one of the entries in the array $U$.  Otherwise, $P_{i,j}$ is part of the code for a base 12 digit for one of the coordinates.  If $i$ is odd and $j$ is even, $P_{i,j}$ is part of the pattern that codes $j_1$, the 12s digit for the $x$ direction.  If $i$ and $j$ are both even, $P_{i,j}$ is part of the pattern that codes $j_0$, the 1s digit for the $x$ direction.  If $i$ is even and $j$ is odd, $P_{i,j}$ is part of the pattern that codes what is in the $4\times 4$ case the only digit for the $y$ coordinate.

Since the pattern is created by meshing together $2\times 2$ arrays, we will introduce some notation to capture how this is done.  For any array $Z$, we will denote by $Z(i,j;m,n)$ the $2\times 2$ array given by
\[
\left[\begin{array}{cc}
Z_{i,j}   & Z_{i,j+n}  \\
Z_{i+m,j} & Z_{i+m,j+n}
\end{array}\right].
\]
Thus, for any nonnegative integers $i$ and $j=12(j_1-1)+(j_0-1)$ with $i\leq 11$ and $j\leq 143$, we have
\[
P(4i+1,4j+1;2,2)=U
\]
\[
P(4i+1,4j+2;2,2)=A_{j_1}
\]
\[
P(4i+2,4j+1;2,2)=(A_i)^t
\]
\[
P(4i+2,4j+2;2,2)=A_{j_0}.
\]

It will also be helpful to introduce names for some of the translations of $U$ and the elements of the sequence $A$.  We will denote by $R$ the set of translations of $U$.  That is,
\[
R=\left\{
\left[\begin{array}{cc}
1 & 0  \\
0 & 0
\end{array}\right],
\left[\begin{array}{cc}
0 & 1  \\
0 & 0
\end{array}\right],
\left[\begin{array}{cc}
0 & 0  \\
1 & 0
\end{array}\right],
\left[\begin{array}{cc}
0 & 0  \\
0 & 1
\end{array}\right]
\right\}.
\]
We can form a sequence of arrays $B$ by taking the second column of each array in sequence $A$ along with the first column of the array that occurs next in the sequence, wrapping around when the end is reached.  That is, define
\[
B_i=
\left[\begin{array}{cc}
(A_i)_{1,2} & (A_{(i\bmod{12}) + 1})_{1,1} \\
(A_i)_{2,2} & (A_{(i\bmod{12}) + 1})_{2,1}
\end{array}\right].
\]
for $i$ from 1 to 12.  This gives
\begin{align*}
B=&\left\{
\left[\begin{array}{cc}
0 & 0 \\
0 & 0
\end{array}\right],
\left[\begin{array}{cc}
1 & 0 \\
1 & 1
\end{array}\right],
\left[\begin{array}{cc}
0 & 0 \\
1 & 1
\end{array}\right],
\left[\begin{array}{cc}
1 & 0 \\
0 & 1
\end{array}\right],
\left[\begin{array}{cc}
1 & 1 \\
1 & 0
\end{array}\right],
\left[\begin{array}{cc}
0 & 1 \\
1 & 1
\end{array}\right],\right.  \\
&\left.\left[\begin{array}{cc}
0 & 1 \\
0 & 1
\end{array}\right]
\left[\begin{array}{cc}
0 & 1 \\
1 & 0
\end{array}\right],
\left[\begin{array}{cc}
1 & 1 \\
0 & 0
\end{array}\right],
\left[\begin{array}{cc}
1 & 1 \\
1 & 1
\end{array}\right],
\left[\begin{array}{cc}
1 & 1 \\
0 & 1
\end{array}\right],
\left[\begin{array}{cc}
1 & 0 \\
1 & 0
\end{array}\right]\right\}.
\end{align*}
Sequences $A$ and $B$ have two important properties.  Every array appears at most once in each sequence and no array in either sequence has exactly one $1$ and three $0$s.

\subsection{Decoding}

Suppose $Y$ is the $4\times 4$ array starting at $P_{4i+m,4j+n}$ with $m,n\in\{1,2,3,4\}$.  The first step in decoding the coordinates $4i+m$ and $4j+n$ is to find $m$ and $n$.  If $m\in\{2,3\}$ and $n\in\{2,3\}$, then
\[
Y(4-m,4-n;2,2)=\left[\begin{array}{cc} 0 & 0 \\ 0 & 1 \end{array}\right].
\]
If $m\in\{2,3\}$ and $n\in\{4,1\}$, then
\[
Y(4-m,2-(n\bmod{4});2,2)=\left[\begin{array}{cc} 0 & 0 \\ 1 & 0 \end{array}\right].
\]
If $m\in\{4,1\}$ and $n\in\{2,3\}$, then
\[
Y(2-(m\bmod{4}),4-n;2,2)=\left[\begin{array}{cc} 0 & 1 \\ 0 & 0 \end{array}\right].
\]
Finally, if $m\in\{4,1\}$ and $n\in\{4,1\}$, then
\[
Y(2-(m\bmod{4}),2-(n\bmod{4});2,2)=\left[\begin{array}{cc} 1 & 0 \\ 0 & 0 \end{array}\right].
\]
Thus, we will have $Y(k,\ell;2,2)\in R$ for some $k,\ell\in\{1,2\}$.  As long as $Y(k,\ell;2,2)\in R$ for only one value of $(k,\ell)$, which is, in fact true, as will be shown later, the cases will not overlap and $m$ and $n$ can be uniquely determined.

Once $m$ and $n$ have been determined, the next step is to find $i$ and $j$.  Suppose $j=12(j_1-1)+(j_0-1)$ where $j_1$ and $j_0$ are nonnegative integers from 1 to 12.  To find $j_0$, the values $Y_{k,\ell}$ for which $m-1+k$ and $n-1+\ell$ are both even can be examined.  That is, let $r$ and $s$ be the least positive integers for which $m-1+r$ and $n-1+s$ are both even and consider the array $Y(r,s;2,2)$.  This array will contain the values that code $j_0$.  There are 4 possible cases.  If $m,n\in\{1,2\}$, then
\[
Y(r,s;2,2)=\left[\begin{array}{cc}
(A_{j_0})_{1,1} & (A_{j_0})_{1,2}  \\
(A_{j_0})_{2,1} & (A_{j_0})_{2,2}
\end{array}\right]=A_{j_0}.
\]
If $m\in\{1,2\}$ and $n\in\{3,4\}$, then
\[
Y(r,s;2,2)=\left[\begin{array}{cc}
(A_{j_0})_{1,2} & (A_{(j_0\bmod 12)+1})_{1,1}  \\
(A_{j_0})_{2,2} & (A_{(j_0\bmod 12)+1})_{2,1}
\end{array}\right]=B_{j_0}.
\]
If $m\in\{3,4\}$ and $n\in\{1,2\}$, then
\[
Y(r,s;2,2)=\left[\begin{array}{cc}
(A_{j_0})_{2,1} & (A_{j_0})_{2,2}  \\
(A_{j_0})_{1,1} & (A_{j_0})_{1,2}
\end{array}\right].
\]
If $m\in\{3,4\}$ and $n\in\{3,4\}$, then
\[
Y(r,s;2,2)=\left[\begin{array}{cc}
(A_{j_0})_{2,2} & (A_{(j_0\bmod 12)+1})_{2,1}  \\
(A_{j_0})_{1,2} & (A_{(j_0\bmod 12)+1})_{1,1}
\end{array}\right].
\]
The last two cases code the same $x$ coordinates as the first two cases respectively, and either of the second two cases can be reduced to one of the first two by flipping the rows of $Y(r,s;2,2)$, so we only have to worry about the first two cases.  In the first case $j_0$ can be found by looking $Y(r,s;2,2)$ up in a table with sequence $A$.  In the second case, $j_0$ can be found by looking up $Y(r,s;2,2)$ in a table with sequence $B$.  In the second case, we will have to keep track of whether $j_0=12$ for the coding of the next digit.

The next digit for the $x$ coordinate can be found in a similar way.  This time, we let $r$ and $s$ be the least positive integers for which $m-1+r$ is odd and $n-1+s$ is even and examine $Y(r,s;2,2)$.  Rather than 4 cases that reduce to 2 by flipping rows, there are now six cases that reduce to 3 by flipping rows.  Specifically, if $m\in\{2,3\}$, we can flip rows to get the same array as one of the following cases in which $m\in\{1,4\}$.  If $m\in\{1,4\}$ and $n\in\{1,2\}$, then
\[
Y(r,s;2,2)=\left[\begin{array}{cc}
(A_{j_1})_{1,1} & (A_{j_1})_{1,2}  \\
(A_{j_1})_{2,1} & (A_{j_1})_{2,2}
\end{array}\right]=A_{j_1}.
\]
If $m\in\{1,4\}$ and $n\in\{3,4\}$ and it is not true that $j_0=12$ and the array coding $j_0$ was part of sequence $B$, then
\[
Y(r,s;2,2)=\left[\begin{array}{cc}
(A_{j_1})_{1,2} & (A_{j_1})_{1,1}  \\
(A_{j_1})_{2,2} & (A_{j_1})_{2,1}
\end{array}\right].
\]
If $m\in\{1,4\}$ and $n\in\{3,4\}$ and we have $j_0=12$ and the array coding $j_0$ was part of sequence $B$, then
\[
Y(r,s;2,2)=\left[\begin{array}{cc}
(A_{j_1})_{1,2} & (A_{(j_1\bmod 12)+1})_{1,1}  \\
(A_{j_1})_{2,2} & (A_{(j_1\bmod 12)+1})_{2,1}
\end{array}\right]=B_{j_1}.
\]
In the second case, $j_1$ can be found by switching the columns of $Y(r,s;2,2)$ to get $A_{j_1}$ and looking up $Y(r,s;2,2)$ in a table with sequence $A$.  In the first and third cases, we can find $j_1$ by looking up $Y(r,s;2,2)$ in a table with sequence $A$ for the first case or $B$ for the third.

In the $4\times 4$ case, the $y$ coordinate is completely determined by the single digit $i$ along with $m$. Once $m$ is known, $i$ can be determined by looking at the values $Y_{k,\ell}$ for which $m-1+k$ is even and $n-1+\ell$ is odd.  This involves finding the least values of $r$ and $s$ for which $m-1+r$ is even and $n-1+s$ is odd and looking at $Y(r,s;2,2)$.  In fact, $i$ can be determined by looking up $Y(r,s;2,2)^t$ in a table with sequence $A$ or $B$ depending on $m$ in a way that is analagous to how the $x$ digits are decoded.

It should now be clear why $Y(k,\ell;2,2)$ can be in $R$ for only one value of $k,\ell\in\{1,2\}$.  Once one value of $k,\ell\in\{1,2\}$ has been found that gives $Y(k,\ell;2,2)\in R$, all other values will give an array consisting of some permutation of the elements in an array in sequence $A$ or sequence $B$.  Since elements of $A$ and $B$ never have exactly one entry equal to 1, no permutation of any of these elements will be in $R$, so no other value of $k,\ell\in\{1,2\}$ will give $Y(k,\ell;2,2)\in R$.

As an example, suppose we are given the $4\times 4$ array $Y$ with the upper left corner at $P_{5,201}$ without knowing the coordinates of the $4\times 4$ array.  This gives
\[
Y=\left[\begin{array}{cccc}
1           & \textcolor{blue}{0} & 0           &  \textcolor{blue}{1}  \\
\textcolor{red}{0} & \textcolor{green}{0} & \textcolor{red}{1} & \textcolor{green}{0}  \\
0           & \textcolor{blue}{1} & 0           & \textcolor{blue}{1}  \\
\textcolor{red}{0} & \textcolor{green}{1} & \textcolor{red}{1} & \textcolor{green}{1}
\end{array}\right].
\]
Since
\[
Y(1,1;2,2)=\left[\begin{array}{cc} 1 & 0 \\ 0 & 0 \end{array}\right],
\]
both digits are congruent to one modulo four, and all of the arrays will be looked up in a table with sequence $A$ without being flipped.  The digits for each coordinate can be found using sequence $A$.  We have
\[
Y(2,2;2,2)=\left[\begin{array}{cc} 0 & 0 \\ 1 & 1 \end{array}\right],
\]
so $j_0=3$,
\[
Y(1,2;2,2)=\left[\begin{array}{cc} 0 & 1 \\ 1 & 1 \end{array}\right],
\]
so $j_1=5$, and
\[
Y(2,1;2,2)=\left[\begin{array}{cc} 0 & 1 \\ 0 & 1 \end{array}\right],
\]
so $i=2$.  To illustrate this more clearly, each of these arrays appears in a different color in the depiction of the array $Y$ above.  This gives $x=((j_1-1)\cdot12+(j_0-1))\cdot 4+1=(4\cdot12+2)\cdot 4+1=201$ and $y=i\cdot 4+1=1\cdot 4+1=5$.

\section*{Acknowledgements}
This work was partially supported by National Science Foundation grant DMS-0451254, which funds a Research Experience for Undergraduates program at Grand Valley State University.

\end{document}